\documentclass[12pt]{iopart}

\usepackage{iopams}
\usepackage[latin1]{inputenc}          
\usepackage{amssymb}                   
\usepackage{graphicx}
\usepackage{alltt}                     
\usepackage{afterpage}
\usepackage{multicol}
\usepackage{epsfig}
\usepackage{fancybox}
\usepackage{shadow}		
\usepackage{epic}			
\usepackage{bar}			
\usepackage{hyperref}

\eqnobysec
\newcommand{\Star}[1]{\,\raisebox{5pt}{$^*$}\hspace{-2.6mm}#1}

\begin{document}
\title{Noether's theorem in classical mechanics revisited}
\author{Rubens M. Marinho Jr.}
\address{Departamento de F\'{\i}sica\\
	Instituto Tecnol\'{o}gico de Aeron\'{a}utica\\
	Brazil}
\ead{marinho@ita.br}
\date{May 1, 2006}
\begin{abstract}
A didatic approach of the Noether's theorem in classical
mechanics is derived and used to obtain the laws of conservation.
\end{abstract}
\maketitle

\section{Introduction}
Noether's\cite{Noether} theorem, presented in 1918, is one of the most
beautiful theorems in physics. It relates symmetries of a theory with its
laws of conservation. Many modern textbooks on quantum field theory present
a pedagogical version of the theorem where its power is demonstrated. The
interested reader is referred to the detailed discussion due to Hill\cite
{Hill}. Despite the great generality of this theorem, few authors present
its version for classical mechanics. See for examle the work of Desloge and
Karcch\cite{Desloge} using
an approach inspired in the work of Lovelock and Hund\cite{Lovelock}.
Several authors demonstrate
Noether's theorem starting from the invariance of the Lagrangian
\cite{Saletan}\cite{Arnold}, but in this case it is not possible to
obtain the energy conservation law in a natural way.

In this article, the theorem is
proved imposing invariance of the action under infinitesimal transformation,
openning the possibility to extend the Noether's theorem in classical 
mechanics to include the energy conservation.

In section 2, the Euler-Lagrange equation is rederived. In section 3
Noether's theorem is proved, in section 4 several applications are
presented and in section 5 the Noether's theorem is extended and
the energy conservation obtained.

\section{The Euler-Lagrange Equations}
\label{sEulerLagrange}
We rederive the Euler-Lagrange equations of motion for sake of completness 
and to introduced notation.

Let us consider a system of particles with $n$ degrees of freedom whose
generalized coordinates and velocities are, respectively, $q$ and $\dot{q},$
characterized by the Lagrangian $L(q,\dot{q},t),$ where $q$ is short hand
for $q_{1}(t),q_{2}(t),\ldots ,q_{n}(t)$, with the dot representing the
total time derivative. When necessary for clarification, the explicit time
dependence will be displayed. This simple system is used in order to place
in evidence the main features of Noether's theorem.

\begin{figure}[h]
	\begin{minipage}[b]{\textwidth}
	 	\setlength{\unitlength}{1cm}
		\vspace{4cm}
		\begin{picture}(5,5)(-2,0) 
			\includegraphics*[width=0.7\textwidth]{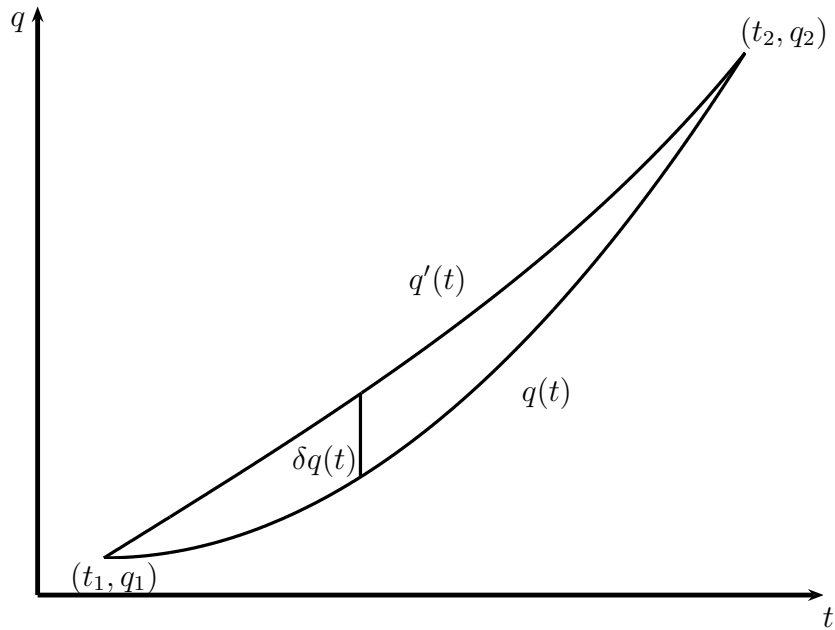}	
			\put(0,0){$t$}
			\put(-10.8,8){$q$}
			\put(-5.5,4.5){$q'(t)$}
			\put(-4,3){$q(t)$}
			\put(-7.05,2.105){$\delta q(t)$}
			\put(-10,0.55){$(t_1,q_1)$}
			\put(-1.1,7.8){$(t_2,q_2)$}
		\end{picture}
	\end{minipage}
	\caption{Varied path of the function $q(t)$}
	\label{fig:variation} 	
\end{figure}

The most general formulation of mechanics is through the principle of least
action or Hamilton's principle\cite{Goldstein,Landau}: {\em the motion of the 
system from fixed time $t_1$ to $t_2$ is such that the action integral 
\begin{equation}
	S=\int_{t_{1}}^{t_{2}}L(q,\dot{q}_{,}t)\rmd t  
	\label{Action}
\end{equation}
is an minimum
\footnote{Actually, in order to obtain the equations of motion we can relax this
restriction imposing only that $S$ be an extremum.} 
for the path $q(t)$ of motion.}
In other words, the variation of the action $\delta S$ is zero for this path
\begin{equation} 
	\delta S =\int_{t_{1}}^{t_{2}}\delta L(q,\dot{q}_{,}t)\rmd t=0.
	\label{deltaS0}
\end{equation}
Using the variation of the Lagrangian in this equation results
\begin{equation}
	\delta S=
		\int_{t_{1}}^{t_{2}}
		\left(
			\frac{\partial L}{\partial q_i}\delta q_i+
			\frac{\partial L}{\partial \dot q_i}\delta \dot q_i
		\right) \rmd t=0,
	\label{deltaS1}
\end{equation}
where the Einstein's summation convention on repeated indices is used. 
The explicit form of the variations in the coordinates and velocities in   (\ref{deltaS1}) are
\begin{equation}
	\delta q(t) = q'(t)-q(t),
	\label{deltaq}
\end{equation}
\begin{equation}
	\delta\dot q(t) =  \frac{\rmd q'(t)}{\rmd t}
-\frac{\rmd q(t)}{\rmd t}=\frac{\rmd }{\rmd t}(q'(t)-q(t))=\frac{\rmd }{\rmd t}\delta q(t),
	\label{deltadq}
\end{equation}
and can be seen in figure \ref{fig:variation}.\\
Integrating the second term of   (\ref{deltaS1}) by parts, using   (\ref{deltadq}) and the condition that the variation of the coordinates at the 
end points of the path $t_1$ and $t_2$ are zero:
\begin{equation}
	\delta q(t_2)=\delta q(t_1)=0,
	\label{endcond}
\end{equation}
gives
\begin{equation}
	\delta S=
		\int_{t_{1}}^{t_{2}}
		\left(
			\frac{\partial L}{\partial q_i}-
			\frac{\rmd }{\rmd t}\frac{\partial L}{\partial \dot q_i}
		\right) \delta q_i \rmd t=0.
\end{equation}
But this is zero for an arbitrary variation $\delta q_i$ only if 
\begin{equation}
	\frac{\rmd }{\rmd t}\frac{\partial L}{\partial \dot q_i} - 
	\frac{\partial L}{\partial q_i} = 0.
	\label{EulerLagrange}
\end{equation}
These are the Euler-Lagrange equations of motion.

The following should also be considered: as is well known, 
the action $S$ is invariant if we replace the Lagrangian, $L$,
of the system by a new Lagrangian, $L'$, 
differing from the old one by the total time derivative of
a function, $g(q,t)$, dependent only on the coordinates and the time. In fact, let us 
consider the new Lagrangian
\begin{equation} L' = L - \frac{\rmd g(q,t)}{\rmd t}. \label{LL} \end{equation}
The new action is
\begin{equation}
	S' = \int_{t_{1}}^{t_{2}}L'(q,\dot{q}_{,}t)\rmd t=\int_{t_{1}}^{t_{2}}\left(L(q,\dot{q}_{,}t)-\frac{\rmd g}{\rmd t}\right)\rmd t
\end{equation}
whose variation is
\begin{equation}
	\delta S' = \int_{t_{1}}^{t_{2}} \delta L'(q,\dot{q}_{,}t)\rmd t
		      =\int_{t_{1}}^{t_{2}}\left(\delta L(q,\dot{q}_{,}t)-\frac{\rmd \delta g}{\rmd t}\right)\rmd t
	\label{deltaSl}
\end{equation}
but using
\begin{equation}
	\delta g(q,t) = \frac{\partial g}{\partial q_i}\delta q_i
\end{equation}
and integrating we obtain
\begin{equation}
	\delta S' = \delta S - \frac{\partial g}{\partial q_i}(\delta q_i(t_2)-\delta q_i(t_1)).
\end{equation}
Using (\ref{endcond}) results in $\delta S' = \delta S$. 

With the help of (\ref{deltaSl}) we conclude that if the
infinitesimal transformation that changes $q$ to $q+\delta q$ is such that the variation
of the Lagrangian can be writenn as a total time derivative of a function $F$:
\begin{equation}
	\delta L = \frac{\rmd \delta g}{\rmd t}=\frac{\rmd F}{\rmd t}
	\label{deltaL}
\end{equation}
then the action $S$ is not affected by the transformation i.e. $\delta S=0$, and $\delta q$ 
is a symmetry of the action.

\section{Noether's Theorem}
If the action of a given system is invariant under the 
infinitesimal transformation that changes $q$ to $q+\delta q$, then, corresponding to
this transformation there exist a law of conservation, and the conserved quantity, $J$, 
can be obtained only from de Lagrangian and the infinitesimal transformation.

In fact, let us supose that the infinitesimal transformation $q'=q+\delta q$ is a symmetry of
the action, then
\begin{equation}
	\delta L = \frac{\partial L}{\partial q_i}\delta q_i+
			\frac{\partial L}{\partial \dot q_i}\delta \dot q_i = \frac{\rmd F}{\rmd t}
\end{equation}
rewritenn this equation using the Euler-Lagrange equations of motion becomes
\begin{equation}
	\frac{\rmd }{\rmd t}\frac{\partial L}{\partial \dot q_i}\delta q_i+
			\frac{\partial L}{\partial \dot q_i}\frac{\rmd }{\rmd t}\delta q_i - \frac{\rmd F}{\rmd t}=0
\end{equation}
but this equation can be put in the form
\begin{equation}
	\frac{\rmd }{\rmd t}\left( \frac{\partial L}{\partial \dot q_i}\delta q_i - F \right)=0.
\end{equation}
The expression inside the parentesis is a conserved quantity named {\it Noether's current}
\begin{equation}
	J = \frac{\partial L}{\partial \dot q_i}\delta q_i - F.
\end{equation}

\section{Several applications of the theorem}
We will examine three important cases of Noether's theorem. The conservation of 
momentum, angular momentum and the moviment of a particle in a 
constant gravitational field. In the
the next section we will extend the Noether's theorem to obtain the 
energy conservation.

\subsection{Momentum conservation}
Momentum conservation is obtained from the freedom we have to 
choose the origin of the system of coordinates.
Let us consider the Lagrangian of a free point particle of mass $m$ moving with 
velocity $\dot\mathbf x$,
\begin{equation}
	L=\frac{1}{2} m \dot x_i\dot x_i.
\end{equation}
Under infinitesimal space translation, 
\begin{eqnarray*}
\left\{
   \begin{array}{rclcl}
      x_i^{\prime } &=&x_{i}+a_i &\rightarrow& \delta x_i = a_i,\\
      \dot x_i' &=& \dot x_i &\rightarrow& \delta\dot x_i = 0,
   \end{array}
\right.
\end{eqnarray*}
the variation of the Lagrangian becomes
\begin{equation}
	\delta L = \frac{\partial L}{\partial x_i}\delta x_i+\frac{\partial L}{\partial\dot x_i}\delta\dot x_i=0.
\end{equation}
The first term is zero because $L$ does not depend on $x_i$ and the second is zero 
because $\delta\dot x_i=0$.
In this case the variation of the Lagrangian can be put in the form of
  (\ref{deltaL}) if we choose $F$ equal a constant $c$.
The Noether's current then results
\begin{equation}
	J = \frac{\partial L}{\partial \dot x_i}\delta x_i - c=m\dot x_i a_i-c=const  \longrightarrow p_i a_i=const.
\end{equation}
As the $a_i$ are arbitrary this is constant only if the momentum $p_i=const$.

\subsection{Angular momentum conservation}
Angular momentum conservation is obtained from the freedom we have to 
choose the orientation of the system of coordinates.
Let us consider the Lagrangian of a free point particle of mass $m$ moving with 
velocity $\dot\mathbf x$ in a plane
\begin{equation}
	L=\frac{1}{2} m \dot x^2+\frac{1}{2} m \dot y^2.
\end{equation}
Under infinitesimal rotation $\theta$, 
\begin{eqnarray*}
\left\{
   \begin{array}{rcrcrcl}
      	x' & = & \cos\theta x+\sin\theta y &=& x + \theta y &\rightarrow& \delta x = \theta y,\\
	y' & =& -\sin\theta x+\cos\theta y &=&   -\theta x +  y &\rightarrow& \delta y = -\theta x
   \end{array}
\right.
\end{eqnarray*}
and 
\begin{eqnarray*}
\left\{
   \begin{array}{rcrcl}
      	\dot x' &=&\dot x+\theta \dot y &\rightarrow& \delta \dot x = \theta \dot y,\\
	\dot y' &=&-\theta \dot x+\dot y &\rightarrow& \delta \dot y = -\theta \dot x,
   \end{array}
\right.
\end{eqnarray*}
the variation of the Lagrangian becomes
\begin{equation}
	\delta L = \frac{\partial L}{\partial x}\delta x+
			\frac{\partial L}{\partial\dot x}\delta\dot x+
			\frac{\partial L}{\partial y}\delta y+
			\frac{\partial L}{\partial\dot y}\delta\dot y=
			m\dot x\theta \dot y+m\dot y(-\theta\dot x)=0.
\end{equation}
Again the variation of the Lagrangian can be put in the form of   (\ref{deltaL}) if we choose $F=c$.
The Noether's current then results
\begin{equation}
	J = 	\frac{\partial L}{\partial \dot x}\delta x+
		\frac{\partial L}{\partial \dot y}\delta y - c= const \longrightarrow
		(xp_y - yp_x)\theta = const.		
\end{equation}
As the infinitesimal angle $\theta$ is arbitrary, the expression inside the parentesis, witch 
is a constant, can be 
recognized as the component $L_z$ of the 
angular momentum.

\subsection{A particle in a gravitational field}
Consider a particle in a constant gravitational field 
described by the Lagrangian
\begin{equation}
	L=\frac{1}{2} m \dot z^2-mgz.
\end{equation}
Under infinitesimal space transformation 
\begin{equation}
	z' =  z+a \rightarrow \delta z = a\rightarrow \delta \dot z = 0.
\end{equation}
The variation of the Lagrangian becomes
\begin{equation}
	\delta L = \frac{\partial L}{\partial z}\delta z+
			\frac{\partial L}{\partial\dot z}\delta\dot z=-mga
\end{equation}
The variation of the Lagrangian can be put in the form of  (\ref{deltaL}) if we choose 
\begin{equation}
	F =-mgat.
\end{equation}
The Noether's current then results
\begin{eqnarray}
	J = 	\frac{\partial L}{\partial \dot z}\delta z - F = 
		m \dot z a +mgat= const.
\end{eqnarray}
In the motion of a particle in a constant gravitational field
the quantity $\dot z+gt$ wich is the initial velocity is conserved.

\section{Extension of the theorem}
With the formalism of the preceding section it is not possible to obtain the 
energy conservation. 
The reason comes from the fact that we have not yet defined what we mean 
by the variation, $\delta t$, in time, necessary to obtain the energy conservation.
In order to define the variation in time let us use another
parametrization for the path described by the particles.
If we use a new parameter $\tau$, the path $q=q(t)$ can be writen
\begin{eqnarray}
	q &=& q(t(\tau))= Q(\tau), \\
	t &=& T(\tau),
\end{eqnarray}
whose variations are
\begin{eqnarray}
	\delta q &=& q'(t(\tau))-q(t(\tau))=Q'(\tau)-Q(\tau)=\delta Q, \\
	\delta t &=& T'(\tau)-T(\tau)=\delta T.
	\label{variations}
\end{eqnarray}
The action  (\ref{Action}) can be writen
\begin{equation}
	S=\int_{t_{1}}^{t_{2}}L(q,\dot{q},t)\rmd t=	
	\int_{\tau_{1}}^{\tau_{2}}L(q,\dot{q},t)\frac{\rmd t}{\rmd \tau}\rmd \tau=
	\int_{\tau_{1}}^{\tau_{2}}{\cal L}(Q,\Star{Q},T,\Star{T})\rmd \tau,
\end{equation}
where ${\cal L}$, $\Star{T}$ and $\Star{Q}$ are defined by the retations
\begin{eqnarray}
	{\cal L} &=& L\Star{T},                                                                       \label{calL}\\ 
	\Star{T} &=& \frac{\rmd t}{\rmd \tau},                                                             \label{TStar}\\
	\dot q &=& \frac{\rmd q}{\rmd \tau}\frac{\rmd \tau}{\rmd t}=\frac{\Star{Q}}{\Star{T}} \label{dotq}.
\end{eqnarray}
The Euler Lagrange equations for this action can be written
\begin{eqnarray}
	\frac{\rmd }{\rmd \tau}\frac{\partial{\cal L}}{\partial\Star{Q}}-\frac{\partial{\cal L}}{\partial Q}=0,\\
	\frac{\rmd }{\rmd \tau}\frac{\partial{\cal L}}{\partial\Star{T}}-\frac{\partial{\cal L}}{\partial T}=0.
	\label{EL}
\end{eqnarray}
If an infinitesimal transformation leaves the action invariant then the variation of the 
Lagrangian can be written as a total time derivative:
\begin{equation}
	\delta {\cal L} = \frac{\partial{\cal L}}{\partial Q}\delta Q+
				 \frac{\partial{\cal L}}{\partial\Star{Q}}\delta\Star{Q}+
				\frac{\partial{\cal L}}{\partial T}\delta T+
				\frac{\partial{\cal L}}{\partial\Star{T}}\delta\Star{T} = \frac{\rmd F}{\rmd \tau}.
\end{equation}
Using the Euler Lagrange equations,  (\ref{EL}), results in
\begin{equation}
	\delta {\cal L} = \frac{\rmd }{\rmd \tau}\left(\frac{\partial{\cal L}}{\partial \Star{Q}}\right)\delta Q+
				 \frac{\partial{\cal L}}{\partial\Star{Q}}\delta\Star{Q}+
				\frac{\rmd }{\rmd \tau}\left(\frac{\partial{\cal L}}{\partial \Star{T}}\right)\delta T+
				\frac{\partial{\cal L}}{\partial\Star{T}}\delta\Star{T} = \frac{\rmd F}{\rmd \tau}
\end{equation}
or
\begin{equation}
	 \frac{\rmd }{\rmd \tau}\left(\frac{\partial{\cal L}}{\partial \Star{Q}}\delta Q+
				      \frac{\partial{\cal L}}{\partial \Star{T}}\delta T
				      - F\right)=0.
\end{equation}
Rewriting in terms of the old variables, using  (\ref{calL},\ref{TStar}) we have
\begin{equation}
	 \frac{\rmd }{\rmd t}\left(\frac{\partial(L\Star{T})}{\partial \Star{Q}}\delta Q+
				      \frac{\partial(L\Star{T})}{\partial \Star{T}}\delta T
				      - F\right)\Star{T}=0.
	\label{antinoether}
\end{equation}
With the help of (\ref{dotq}) the following relations holds
\begin{eqnarray}
	\frac{\partial L}{\partial \Star{Q}}\Star{T}  = 
	\frac{\partial L}{\partial \dot q}\frac{\partial\dot q}{\partial \Star{Q}}\Star{T} = 
	\frac{\partial L}{\partial \dot q},\\
	\frac{\partial L}{\partial \Star{T}}\Star{T}  =
	\frac{\partial L}{\partial \dot q}\frac{\partial\dot q}{\partial \Star{T}}\Star{T} =
	\frac{\partial L}{\partial \dot q}\left(-\frac{\Star{Q}}{\Star{T}}\right)=-\frac{\partial L}{\partial \dot q}\dot q,
\end{eqnarray}
whose substitution in  (\ref{antinoether}), remembering that $Q$ and $T$
are independent variables in the new Lagrangian $\cal L$, gives
\begin{equation}
	\frac{\rmd }{\rmd t}\left[\frac{\partial L}{\partial\dot q}\delta q+
	\left(-\frac{\partial L}{\partial\dot q}\dot q+L\right)\delta t-F\right]=0.
\end{equation}
Recognizing the term inside de brackets as minus the Hamiltonian results for the Noether's current
\begin{equation}
	J = \frac{\partial L}{\partial\dot q}\delta q - H\delta t -F = const.
	\label{ExtendedJ}
\end{equation}

\subsection{Conservation of energy}
Energy conservation is based on the freedom we have to choose the origin of the time. 
Let us cosider, as in general is the case, a Lagrangian $L(q,\dot q, t)$ not dependent explicitly on time.
Under an infinitesimal time translation
\begin{eqnarray*}
\left\{
   \begin{array}{ccl}
      q' &=&q \longrightarrow \delta q = 0,\\
      t' &=&t+\epsilon \longrightarrow \delta t = \epsilon,
   \end{array}
\right.
\end{eqnarray*}
the variation of the Lagrangian, 
\begin{equation}
	\delta L = \frac{\partial L}{\partial q_i}\delta q_i+
			\frac{\partial L}{\partial \dot q_i}\delta \dot q_i+\frac{\partial L}{\partial t}\delta t=0,
\end{equation}
so, again, the variation of the Lagrangian can be put in the form of
  (\ref{deltaL}) if we choose $F=c$.
The conserved current
\begin{equation}
	J = \frac{\partial L}{\partial \dot q_i}\delta q_i-H\delta t - c=const \longrightarrow H\epsilon = const.
\end{equation}
As $\epsilon$ is an arbitrary quantity, in order that $J$ be constant $H$ must be constant and can be recognized as the energy of the system. In other words, if the
Lagrangian is invariant under time translation,  then the energy is conserved.

\section{Conclusion}
The aim of this work was to present in  a didatic way 
the Noether's theorem in the scope of Classical Mechanics. 
This theorem is not so important in Classical Mechanics as it is
in Field Theory, this is the reason that its didatic presentation
generaly comes in 
textbooks of Field Theory.  
Even in books that treat the theorem in Classical Mechanics,
such as Saletan\cite{Saletan} and Arnold\cite{Arnold},
do not extend the theorem to include the case of energy conservation.

We hope that this work can bring the main ideas of this theorem 
to undergraduate students in a clear way.

\end{document}